\documentclass[aps,pra,twocolumn,notitlepage,showpacs,superscriptaddress,nofootinbib,floatfix]{revtex4-1}
\usepackage{graphicx,amsfonts,amssymb,amsmath,enumerate,soul,pbox,multirow,xcolor}
\usepackage[normalem]{ulem}
\usepackage[english]{babel}
\usepackage{url}
\usepackage{breakurl}
\usepackage[breaklinks]{hyperref}

\newif\ifhyper
\hypertrue
\ifhyper
\hypersetup{
   citecolor = {green},
   colorlinks = {true}, % false
   urlcolor = {blue} % magenta
}
\fi

\newcommand{\beq}{\begin{equation}}
\newcommand{\eeq}{\end{equation}}
\newcommand{\beqa}{\begin{eqnarray}}
\newcommand{\eeqa}{\end{eqnarray}}
\newcommand{\ket} [1] {\vert #1 \rangle}

\newcommand{\braket}[2]{\langle #1 | #2 \rangle}

\def\ket#1{\vert#1\rangle}

\def\Longarrow{\protect\@lra}
\def\@lra{\relbar\joinrel\relbar\joinrel\relbar\joinrel%
          \relbar\joinrel\rightarrow}

\begin{document}

\title{Quantum computing for finance: overview and prospects}

\author{Rom\'an Or\'us}
\affiliation{Institute of Physics, Johannes Gutenberg University, 55099 Mainz, Germany}
\affiliation{Donostia International Physics Center, Paseo Manuel de Lardizabal 4, E-20018 San Sebasti\'an, Spain}
\affiliation{Ikerbasque Foundation for Science, Maria Diaz de Haro 3, E-48013 Bilbao, Spain}
\affiliation{Quantum for Quants Commission, Quantum World Association, Barcelona, Spain}

\author{Samuel Mugel}
\affiliation{Quantum for Quants Commission, Quantum World Association, Barcelona, Spain}
\affiliation{The Quantum Revolution Fund, Carrer de l'Escar 26, 08039 Barcelona, Spain}

\author{Enrique Lizaso}
\affiliation{Quantum for Quants Commission, Quantum World Association, Barcelona, Spain}

\begin{abstract}

We discuss how quantum computation can be applied to financial problems, providing an overview of current approaches and potential prospects. We review quantum optimization algorithms, and expose how quantum annealers can be used to optimize portfolios,  find arbitrage opportunities, and  perform credit scoring. We also discuss deep-learning in finance, and suggestions to improve these methods through quantum machine learning. Finally, we consider quantum amplitude estimation, and how it can result in a quantum speed-up for Monte Carlo sampling. This has direct applications to many current financial methods, including pricing of derivatives and risk analysis. Perspectives are also discussed.

\end{abstract}

\maketitle

\section{Introduction}

The 50's saw the rise of digital, or classical computing, which has known tremendous success, owing to the immense computational power these machines have given us. As of the 80's, scientists started considering approaching numerical calculations from an entirely new perspective: using the intrinsic, quantum mechanical properties of matter to solve difficult calculations \cite{Nielsen2010}. This marked the conceptual birth of quantum computing.

Relative to classical information processing, quantum computation holds the promise of highly efficient algorithms, providing exponential speedups for some technologically important problems \cite{qft}. While only small quantum processors are currently available, there are tremendous expectations for this technology, mainly due to the widespread belief that quantum computing will experience an enormous growth rate in the near future.

The race to the quantum computer is largely motivated by the shear amount of technological disruption this machine is expected to bring\footnote{See, e.g., \href{https://en.wikipedia.org/wiki/Quantum$\_$computing}{https://en.wikipedia.org/wiki/Quantum$\_$computing}}. Of crucial importance, we can expect our approach to \emph{finance} to be completely transformed. Broadly speaking, finance can be defined as the science of money management, a discipline almost as old as civilization itself. Among the huge variety of problems finance attempts to address, we find stock markets prediction, portfolio optimization, and fraud detection. 

The idea of applying quantum mechanics to finance is not a new one: some well-known financial problems can be directly expressed in a quantum-mechanical form. As an example, the Black-Scholes-Merton formula  \cite{bsm, bsm2} can be mapped to the Schr$\ddot{{\rm o}}$dinger equation, modeling the arbitrage relationships that led to its formulation. Even the entire financial market can be modeled as a quantum process, where quantities that are important to finance, such as the covariance matrix, emerge naturally \cite{Haven2002,baaquie2007}.

In this paper, we provide a first overview of different fields within finance which could benefit from a computational speedup using quantum computers. As we describe bellow, this speedup could manifest itself in a number of different ways, each of which could imply gargantuan savings for governments, financial institutions, and individuals.

Many problems in finance can be expressed as optimization problems. These are tasks which are particularly hard for classical computers, but find a natural formulation using quantum optimization methods \cite{aqc}. In recent years, this field has known a tremendous growth, partly due to the commercial availability of quantum annealers.

Another way to approach financial problems is to search for patterns in past data. This is a natural way to consider economic forecasting problems, an area where machine learning methods have proved to be extremely successful. The computational cost of these approaches, however, is often prohibitive. In recent years, there has been an impressive effort to develop quantum machine learning algorithms \cite{Biamonte2016}, which many hope will provide the tools to satisfy our growing data requirements.

Moreover, the behavior of some financial systems can be predicted by applying Monte Carlo methods. The act of sampling a distribution function can limit the speed, and hence the applicability, of the algorithm. In a series of recent papers, it was suggested this task could be done efficiently by sampling a quantum system \cite{qaest, Montanaro2015, Rebentrost2018}.

{As a word of caution, the financial models we study here have been deemed trustworthy by the community due to their ability to replicate the past behavior of financial markets. It is important to realize that even widely accepted models may make erroneous predictions in qualitatively new situations. One spectacular example is the crash of 2008, caused to a large extent by extrapolating the past low-risk performance of mortgage-based assets to the qualitatively different situation created by the proliferation of subprime mortgages. While quantum computing provides powerful computational tools, whether or not it can predict this type of events remains to be proven.}

The structure of this paper is as follows: in Sec.\ \ref{sec2}, we provide a basic background on financial problems, common algorithms in quantitative finance, and quantum computing. In Sec.\ \ref{sec3}, we examine applications of quantum optimization to finance. In Sec.\ \ref{sec4}, we introduce quantum machine learning (QML), and describe situations where it can be of relevance to financial problems. Financial applications of quantum amplitude estimation to Monte Carlo sampling are detailed in Sec.\ \ref{sec5}.  Finally, in Sec.\ \ref{sec7}, we conclude and discuss the perspectives.  

\section{Background}
\label{sec2}

We will begin by providing some basic background on relevant concepts in finance and quantum computing. This review does not aim to be exhaustive. We refer the readers interested in more in-depth discussions and derivations to the excellent books by Wilmott on quantitative finance \cite{Wilmott} and by Nielsen and Chuang on quantum computing \cite{Nielsen2010}. 

\subsection{Some core problems in finance} 
\label{FinanceCoreProblems}

In its deepest nature, finance deals with the uncertainty in the future behavior of an asset, and the prices and \emph{returns} (profits or losses) it may have in the future. The concept of \emph{risk} quantifies the possibility that the \emph{actual} return of the asset may differ from the \emph{expected} return (that the investor originally had in mind). The measure of \emph{risk} depends on the distribution of returns. This defines \emph{volatility}: the degree of variation of a trading price series over time, as measured by the standard deviation of logarithmic returns. 

\begin{table}
	\centering
	\begin{tabular}{||c|c||} 
	\hline 
 Question & ~~~Broad approach solution~~~  \\
         \hline 
         \hline
         & \\
        \pbox{4cm} {\emph{Which assets should be included in an optimum portfolio? 
        How should the composition of the
         portfolio change according to what 
         happens in the market?}} & Optimization models \\
         & \\
         \pbox{4cm}{\emph{How to detect opportunities 
         in the different assets in the market,
         and take profit by trading with them?}} & \pbox{4cm}{Machine learning methods, including 
         neural networks and deep learning} \\
         & \\
         \pbox{4cm}{\emph{How to estimate the risk and return 
         of a portfolio, or even a company?}} & Monte Carlo-based methods \\
         & \\
         \hline
         \end{tabular}
         \caption{Financial problems addressed in this paper, and possible approaches.} 
         \label{table0} 
\end{table}

To lower the risk, a possibility is to analyze the behavior of the asset, linking it to market information. This is the realm of \emph{financial prediction}, plagued with problems of great practical and theoretical interest. Artificial intelligence techniques are particularly successful at solving this type of problems.

We may mitigate the risk of holding asset $A$ by carefully selecting other additional assets to invest in, either with anticorrelated returns (\emph{hedging}), or uncorrelated ones (\emph{diversifying}). These concepts lead to the definition of \emph{optimal portfolio}: for a given risk, there is one portfolio that maximizes the return. Conversely, for a given return, there is one portfolio of assets that minimizes the risk. An interesting problem arises: how to construct this portfolio, and how to modify it depending on the conditions of the market?

Due to our incomplete knowledge of the market, it is generally convenient to think of assets and portfolios as intrinsically random systems. This randomness is a source of risk which can be extremely difficult to estimate. This is for instance the case of \emph{options}, which are a special case of derivative security. Options' payoffs depend on the value of other assets (hence the name of derivatives) in complex ways. In its essence, it is an agreement which grants one party the right, but not the obligation, to buy or sell an asset at a pre-agreed price. The problem of what an option is worth can be solved, in simple instances, by closed formulas \cite{bsm, bsm2}, but in general requires numerical simulation methods (such as Monte Carlo).

\subsection{Relevant approaches in quantitative finance}
\label{secIIB}

In this section, we will describe three computational approaches to financial problems. These are the focus this paper, and are summarized in Table \ref{table0}. 

Dynamic Portfolio Selection is an example of \emph{optimization}: given a selection of assets $i$, with $i \in [1, n]$, the goal is to maximize the final profits at time $t=T$. Let $x_{i,t}$ be the amount of money we choose invest in asset $i$ at time $t\in [0,T-1]$, and let $R_{i,t+1}$ be the returns resulting from this decision. Given a return $R_{i,t+1}$, the decision to invest $x_{i,t+1}$ at the next time step is usually computed iteratively. This can be done using classic linear programming in simple cases, but for complex problems calls for more elaborate methods such as simulated annealing. The dynamic stochastic problem is sketched in Table \ref{tabprob}.

\begin{table}
	\centering
	\begin{tabular}{|l l|} 
         \hline
       \multirow{3}{*}{Problem}
       & Find: $\max_{x,w}$ $E(U(w_T))$, with: \\
       & $E=$ expectation function \\
       & $U=$ utility function \\
       & \\
       \multirow{6}{*}{Constraints~}
       & $\forall t \in [0, T-1]$:\\
       & $w_{t+1} = \sum_{i=1}^n \left( 1 + R_{i, t+1} \right) \cdot x_{i,t}$\\
       & $\sum_{i = 1}^n x_{it} = w_t$ \\
       & $w_t \ge 0$ \\
       & $w_t$ are random, except $w_0$ \\
       & $x_t$ are random, except $x_0$ \\
       \hline
        \end{tabular}
        \caption{Dynamic stochastic problem of portfolio optimization.} 
         \label{tabprob} 
\end{table}

% \emph{Machine learning methods} train a model using a dataset; the trained model is then used to predict the behavior of new data points. The basic problem here is to find a predictor of an output $Y$ given an input $X$. The machine attempts to learn the mapping $Y = F(X)$, where $X = (X_1, \ldots , X_p)$, and a predictor is denoted by $Y^{\hat{~}}(X) := F(X)$. The output can be continuous, discrete (as in classification problems) or mixed. 

Given a dataset, a model can be trained to identify patterns in the data. This model can then be used to predict the behavior of new data points. This is the basic idea behind \emph{machine learning}. Specifically, given a dataset $\vec{X}$ which produce outputs $\vec{Y}$, the model attempts to learn the mapping $\vec{Y} = F(\vec{X})$. Some forms of machine learning, such as deep learning in neural networks, are distinguished by the use of sequential levels of processing, passing learned features of data through different layers of abstraction. The computing burden lies in training the model, not in making predictions. Heaton et al.\ \cite{Heaton} offers a detailed description of deep learning in finance.

\emph{Monte Carlo methods} are a powerful statistical sampling method. These are extremely useful for modeling complex systems, such as the value of an asset $S$ at time $t$. Under the assumption $S$ can be modeled using a risk-neutral random walk, its evolution is given by:
\beq
dS_t = S_t \alpha dt + S_t \sigma dW_t, 
\eeq
with $\alpha$ the drift, $\sigma$ the volatility, and $dW_{t}^2 = dt$ {(i.e., a Wiener process, or Brownian motion).}  Simulating this random walk may be done by using $\delta S_t = \alpha S_t \delta t + S_t \sigma \left( \delta t \right)^{1/2} \phi$, where $\phi$ is a sample from a normal distribution. In the case of a lognormal random walk, there is a closed formula to calculate the value of the asset at time $t+1$: 
\beq
S_{t+1} = S_t e^{ \left( \left( \alpha - \sigma^2/2 \right)\delta t + \sigma \left( \delta t \right)^{1/2} \phi \right) }.
\eeq
Monte Carlo methods can usually be implemented efficiently, but require many runs to provide an accurate estimation of the expected return and its distribution. {Moreover, this type of modeling of financial markets shows less prediction accuracy for short times, due to the assumption that the drift and volatility parameters are constant. The accuracy can however be improved by further modeling these parameters as stochastic functions of time as well as of other macroeconomic factors.}

While these three approaches have proved to be very successful when applied to financial problems, they invariably require colossal computational power to accurately describe the system, a problem which worsens as the amount of data we gather increases. In this situation, faster means to run these algorithms would be highly disruptive to the industry. 

\subsection{Some basics on quantum computing}

A qubit is the minimum amount of processable information in quantum computing: a two-dimensional quantum-mechanical system, which encodes the classical bits of information $0$ and $1$ in its basis states: $\ket{0}$ and $\ket{1}$. This system can be in a \emph{superposition} of states $\ket{0}$ and $\ket{1}$. This is the first crucial property of quantum systems: they can be simultaneously in all of the system's states at once. It is this property which allows quantum computers to perform parallel computations on a massive scale.

Given a two qubit system, it is possible that the state of each qubit cannot be described individually. Mathematically, the state cannot be factorized as the tensor product of two separate states. When this is true, we say the states are \emph{entangled}. While systems which do not display {too much} entanglement can be described efficiently using classical computational methods such as tensor networks  \cite{Orus2014}, {certain} {classes of highly-entangled} entangled systems are very difficult for classical computers to model. Consequently, for a quantum algorithm to {exceed the capabilities of the best possible classical algorithm, it {necessarily must exploit large amounts of} entanglement.}
% provide an advantage over the best equivalent classical algorithm, it has to use entanglement.
Entanglement also finds applications outside of quantum computing, for instance in quantum cryptography \cite{Eckert91}, quantum teleportation \cite{teleport}, and quantum sensors.

Another fundamental difference between classical and quantum computing is in the basic set of operations. Classical computing is based on binary operations, such as the $NOT$ and $AND$ gates. These operations are \emph{universal}: any other boolean operation can be replicated using $NOT$s and $AND$s. They are also non-reversible: given the result of an $AND$ gate, I cannot deduce the input variables. By contrast, quantum evolution is reversible, as dictated by the Schr$\ddot{{\rm o}}$dinger equation. Events which destroy reversibility, such as measurements, lead to a loss of quantum behavior. To have a quantum gain, it is important to only use reversible,  unitary gates \cite{reversible}. It can be shown that a small set of these gates are also universal.

In general, a quantum algorithm is a sequence of five steps:
\begin{enumerate}
 \item Encode the input data into the state of a set of qubits.
 \item Bring the qubits into superposition over many states (i.e., use \emph{quantum superposition}).
 \item Apply an algorithm (or oracle) \emph{simultaneously} to all the states (i.e., use \emph{quantum entanglement} amongst the qubits); at the end of this step, one of these states holds the correct answer.
 \item Amplify the probability of measuring the correct state (i.e., use \emph{quantum interference}).
 \item Measure one or more qubits.
\end{enumerate}
According to quantum mechanics, the result of the measurement is random. We want to engineer the algorithm so that the most probable answer is interpretable as a classical result which encodes the solution to our problem.

\subsection{The case for quantum computing in finance} 
\label{QCforfinance}

Various quantum algorithms offer substantial speedups relative to classical algorithms. This means that, when the number of classical bits needed to specify the input data is increased, the number of operations needed to run the quantum algorithm increases slower than the best known classical alternative. In this section, we outline some quantum algorithms which are potentially applicable to financial problems. In what follows, $N$ specifies the number of possible inputs (aka the size of the problem), which can be codified using $\log N$ classical bits. 

One cannot talk about the great breakthroughs that started the field of quantum computing without mentioning Grover's algorithm \cite{Grover1996}, which finds a particular register in an unordered database in $\mathcal{O}(\sqrt{N})$ steps. By contrast, the best classical algorithm requires $\mathcal{O}(N/2)$ steps. This algorithm can be adapted to solve optimization problems, such as finding a Minimum Spanning Tree \cite{Aghaei2008}, maximizing flow-like variables \cite{Ambainis2006}, and implementing Monte Carlo methods \cite{qaest}. Alternatively, the Quantum Approximate Optimization Algorithm (QAOA) finds a ``good solution" (i.e: one with a minimum quality) to an optimization problem in a polynomial time \cite{qaoa}. This requires exponential time on a classical computer.

Optimization techniques are also applicable to machine learning algorithms. Indeed, training can be considered as a special case of optimization for neural networks. Machine learning also make frequent use of Fourier transforms. Here again quantum computers can result in a substantial speedup: while classical Fast Fourier Transform runs in $\mathcal{O}(N \log N)$ steps, the Quantum Fourier Transform (QFT) has complexity $\mathcal{O}((\log N)^2)$ \cite{qft}. QFT can be useful for some artificial intelligence methods, such as Quantum Principal Component Analysis (PCA) \cite{Lloyd2014} and Quantum Support Vector Machines (SVM) \cite{Rebentrost2014}.

\begin{table}
	\centering
	\begin{tabular}{||c|c||} 
	\hline 
	Method & ~~~Speedup~~~  \\
         \hline 
         \hline
         Bayesian inference \cite{tab1, tab11} & $\mathcal{O}(\sqrt{N})$ \\
         Online perceptron \cite{tab2} & $\mathcal{O}(\sqrt{N})$ \\
         Least-squares fitting \cite{Wiebe2012} & $\mathcal{O}(\log N)$ \\
         Classical Boltzmann machine \cite{tab4} & $\mathcal{O}(\sqrt{N})$ \\
         ~~~~Quantum Boltzmann machine \cite{tab5, tab55} ~~~~& $\mathcal{O}(\log N)$ \\
         Quantum PCA \cite{Lloyd2014} & $\mathcal{O}(\log N)$ \\
         Quantum support vector machine \cite{Rebentrost2014} & $\mathcal{O}(\log N)$ \\
         Quantum reinforcement learning \cite{tab8} & $\mathcal{O}(\sqrt{N})$ \\
         \hline
         \end{tabular}
\caption{Speedups offered by several QML subroutines, as explained in the table of Box 1 in Ref.\ \cite{Biamonte2016}. Here we adopt the same notation as in that reference: $\mathcal{O}(\sqrt{N})$ is a square-root speedup, and $\mathcal{O}(\log N)$ is an exponential speedup. For more details on their implementations, see Ref.\ \cite{Biamonte2016}.}
\label{tab1}
\end{table}

The Harrow, Hassidim and Lloyd (HHL) algorithm, which solves linear systems of equations, shows an exponential improvement relative to the best classical alternative \cite{Harrow2009}. In recent years, this algorithm has caused a lot of excitement in the field, mainly due to its wide applicability. Because matrix operations are central to many machine learning algorithms, such as pattern recognition, many QML methods make use of the HHL algorithm. A summary of speedups available for QML algorithms is provided in Table \ref{tab1}.

\subsection{Currently available quantum hardware}

It is possible to classify the quantum computing hardware community in two main families. On one hand, quantum computers based on the quantum gate model and quantum circuits, which are the most similar to our current classical computers based on logical gates. In terms of the number of qubits (for the gate model architecture), Google is the current record holder with a 72 qubit quantum processor. There exist a number of different strategies for implementing physical qubits. The main companies currently developing general-purpose quantum processors (by strategy), are Alibaba, IBM, Google, and Rigetti (using superconducting qubits), IonQ (using trapped ion qubits), Xanadu (developing  a photonic quantum computer), and Microsoft (using topological qubits).

The other great family of quantum computers are \emph{quantum annealers}. These computers are designed with the purpose of finding local minima in combinatorial optimization problems. Some experimental quantum annealers are already commercially available, the most prominent example being the D-Wave processor, which sports over 2000 superconducting qubits. This machine has been heavily tested in laboratories and companies worldwide, including Google, LANL, Texas A\&M, USC, and more. Other small-scale quantum annealers are already pursued by initiatives and start-ups, such as Qilimanjaro (which also use superconducting qubits), and NTT (developing a photonic quantum annealer). Under ideal circumstances, these quantum computers are as powerful as those based on the quantum circuit model \cite{aqcuniversal}.

The list established in this section is by no means exhaustive. We refer the reader to Ref.\ \cite{QubitCount} for a more complete list of functioning quantum computers.

% Some issues need to be addressed to construct a fully operable quantum computer. One of the most important problems is decoherence, i.e: uncontrolled interactions between the system and its environment. This leads to a loss of quantum behavior in the quantum processor. Thus, the run-time of quantum algorithms are limited by the decoherence time. If the error rate is small enough, this problem can be addressed by using of quantum error-correcting techniques. Depending on the machine's properties, implementing error correcting algorithms can require as much as a thousand physical qubits to operate one logical qubit.

{
\subsection{Challenges for quantum computing}

We caution the reader that constructing a quantum computer which is capable of outperforming classical computers is a truly formidable task, and potentially one of the great challenges of the century. Before we reach this level, a number of critical issues will have to be dealt with.

One of the most important problems is decoherence, i.e: uncontrolled interactions between the system and its environment. This leads to a loss of quantum behavior in the quantum processor, killing any advantage that a quantum algorithm could provide. The decoherence time therefore sets a hard limit on the number of operations we can perform in our quantum algorithm. An important hardware challenge is the design of higher fidelity qubits. {As such, qubits must be considered as embedded on an open environment, for which classical simulation software packages may be useful \cite{QuTiP, QuTiP2}.}

It is possible to correct for decoherence using error-correction algorithms. This can be done by encoding the quantum state, with redundancy, over many qubits, and is only possible when the error rate {of individual quantum gates} is sufficiently small. With these, we can fully build quantum algorithms which run for longer than the decoherence time. A huge obstacle we are facing is that operating a single fault-tolerant qubit can require many thousands of physical qubits. In a recent study, it was estimated that quantum computing could achieve a significant speedup (in absolute time), but this advantage vanished when the classical processing required to implement error-correction schemes was taken into account \cite{Campbell2018}. Another important challenge is therefore the development of new error-correction schemes with more reasonable requirements.

In view of these obstacles, many researchers have turned to algorithms for so called Noisy Intermediate-Scale Quantum (NISQ) processors. These are built to run on faulty quantum computers and produce good results despite decoherence. It is an extremely exciting branch of quantum computing, both highly versatile and a prime candidate to be the first to achieve quantum supremacy \cite{Katzgraber2018,Iyer2018,Benedetti2018,Santra2018,McClean2018}. Being an extremely new direction of study, we lack an important algorithm library for NISQ machines. This is another great software challenge: to develop new algorithms to make near term quantum computers applicable to real world problems.

Quantum computing has been suggested as a solution to many computationally demanding problems, especially in machine learning, which require processing vast amounts of data. At present, we do not have a quantum RAM (qRAM) capable of efficiently encoding this information as a quantum state, and reliably storing it for extended periods of time. This is among the largest hardware challenges for quantum computing.
}

\section{Quantum optimization}
\label{sec3}

Optimization problems are at the core of many financial problems. This is the case, for instance, of portfolio optimization, which we will discuss in the following \cite{Rosenberg2016}. Because it is an NP-Hard problem, it is extremely difficult, if not impossible, for classical computers to efficiently determine the best choice of portfolio. There are a number of different ways to implement \emph{quantum optimization} algorithms on a quantum computer, the most prominent of which is quantum annealing.

At the heart of quantum optimization algorithms is a method known as adiabatic quantum computation \cite{aqc}, which we will describe in the following. First, we must map the optimization problem to the physical problem of finding the ground state of a Hamiltonian $H_P$, which encodes the cost function to be minimized. We prepare the system in the ground state of an initial Hamiltonian $H_0$, chosen because its ground state is known and easy to prepare. We then adiabatically deform $H_0$ to $H_P$ over a long time $T$. The adiabatic theorem states that a system initiated in its ground state will always remain close to its instantaneous ground state, provided its lowest energy levels are non-degenerate and that the evolution is slow \cite{Kato1950}. In practice, we usually choose $T = \mathcal{O}(\Delta^2)$ with $\Delta$ the minimum energy difference between the instantaneous ground and first exited state. When this is true, measuring the state of the system at the end of the evolution has a a high probability of returning the ground state of $H_P$. This is a universal model of quantum computation \cite{aqcuniversal}, which means that it can in principle perform any quantum algorithm. Furthermore, it is an extremely general model, as it can be modified to add intermediate Hamiltonians \cite{aqcinterm} and can be made to fulfill local adiabatic conditions \cite{aqclocal}.

Quantum annealing is the physical process of implementing an adiabatic quantum computation. This process is similar to classical or simulated annealing, where thermal fluctuations allow the system to jump between different local minima in the energy landscape. As the temperature is lowered, the probability of moving to a worse solution tends to zero. In quantum annealing, these jumps are driven by quantum tunneling events. This process explores the landscape of local minima more efficiently than thermal noise, especially when the energy barriers are tall and narrow\footnote{See, e.g., \href{https://en.wikipedia.org/wiki/Quantum_annealing}{https://en.wikipedia.org/wiki/Quantum$\_$annealing}}.

In practice, it is difficult to fulfill the conditions required for adiabatic quantum computing in a quantum annealing process. It can be difficult, for instance, to guarantee that the system's evolution is fully adiabatic, or that the system is initiated in the true ground state of the initial Hamiltonian. {Quantum annealing is therefore an approximate realization of adiabatic computing. It was shown by Zagoskin that approximate adiabatic computing can find a solution which is close to optimal -- a problem which is also known to be NP-hard -- in polynomial time \cite{Zagoskin2011}.}

Let us now focus on three case-examples where quantum optimization has been used successfully in practical financial problems. While these results are proof-of-principles, they demonstrate that, in the near future, quantum annealers will have practical value with regards to problems in finance.

\subsection{Optimal trading trajectory}

Let us consider the problem of dynamic portfolio optimization, described in Sec.\ \ref{secIIB}. Our aim is to find the optimal trajectory in the portfolio space, while taking into account transaction costs and market impact. 

It was suggested in Ref.\ \cite{delprado} that the discrete multi-period version of this problem was amenable to quantum annealers. This idea was implemented on a D-Wave quantum processor in Ref.\ \cite{Rosenberg2016}. The cost function which was optimized was the return:
\beq
w = \sum_{t = 1}^T \left( \mu_t^T w_t - \frac{\gamma}{2} w_t^T \Sigma_t w_t - \Delta w_t^T \Lambda_t \Delta w_t + \Delta w_t^T \Lambda_t' w_t \right), 
\eeq
with $\mu$ the forecast returns, $w$ the holdings, $\Sigma$ the forecast covariance tensor, and $\gamma$ the risk aversion. The third and fourth terms represent different contributions to transaction costs (see Ref.\ \cite{Rosenberg2016} for details). The overall return must be optimized under the constraint that the sum of holdings is equal to $K$ at all time steps, 
\beq
\sum_{n = 1}^N w_{nt} = K ~~~ \forall t, 
\eeq
and that the sum of the maximum allowed holdings of each asset at any time be at most $K'$, 
\beq
w_{nt}\le K' ~~~ \forall t , ~ \forall n . 
\eeq

This problem was solved on two D-Wave chips with 512 and 1152 qubits \cite{Rosenberg2016}. While only small instances were implemented, the performance of the quantum annealers was similar to that of classical hardware. These experiments proved that this problem can be solved on the D-Wave machine with a high success rate. It was also observed that a proper fine-tuning of the D-Wave machines allowed for important improvements in success rates. The prospect is that future versions of the D-Wave chip should soon be able to handle much bigger instances of the problem, eventually overtaking classical methods. 

\subsection{Optimal arbitrage opportunities}

The concept of arbitrage is the idea of making profit from differing prices in the same asset in different markets. For instance, we could change euros for dollars, then to yens, and then back to euros, and make a small profit in the process. This is cross-currency arbitrage. There exist several classical algorithms which are able to efficiently determine if, for a given set of assets and transaction costs, there exists a cycle that provides a positive return. The problem of determining the optimal arbitrage opportunity, however, is NP-Hard. 

As shown by Rosenberg, optimal arbitrage opportunities can be detected using a quantum annealer \cite{Rosenberg}. Essentially, one starts by constructing a directed graph, where the nodes $i$ represent the assets and the directed edges are weighted with the conversion rate $c_{ij}$. In general, conversion rates are not symmetric, i.e: $c_{ij} \neq c_{ji}$, and transaction costs are assumed to be included in the variable. The optimization problem can be solved by finding the most profitable cycle in this directed graph. 

The problem can be conveniently recast in terms of boolean variables $x_{ij}$, which are $1$ if the link $\{ij \}$ belongs to the cycle, and $0$ otherwise. The figure of merit to be minimized is:
\beqa
w = & & \sum_{(i,j) \in E} x_{ij} \log c_{ij} \nonumber \\
&-& M_1 \sum_{i \in V} \left( \sum_{j, (i,j) \in E} x_{ij} - \sum_{j, (j,i) \in E} x_{ji} \right)^2 \nonumber \\
&-& M_2 \sum_{i \in V} \sum_{j, (i,j) \in E} x_{ij} \left(\sum_{j,(i,j) \in E} x_{ij} - 1 \right). 
\eeqa
In this equation, $E$ are the edges and $V$ the vertices of the graph. The first term represents the logarithm of the cost of the cycle. The second term represents a flow constraint that forces the solution to be a cycle. The third term constrains $x_{ij}$ to be either $0$ or $1$, so that cycles can only go once through any given asset. $M_1$ and $M_2$ are adjustable penalty parameters.

Written in this way, the problem has been boiled down to a quadratic unconstrained binary optimization (QUBO) problem, which is amenable to quantum annealers. These results were implemented on the D-Wave 2X quantum annealer, for a small-size example with five assets. It was found that the quantum annealer produced the same optimal solutions as an exhaustive classical solver \cite{Rosenberg}. The authors extended this study by introducing risk variables, and accounted for cases in which one asset can be bought multiple times.

\subsection{Optimal feature selection in credit scoring} 
\label{optcred} 

It is essential for banks and other financial institutions which specialize in lending money to estimate the level of risk associated with a loan, i.e: if the borrower is likely to default on his payments. This is \emph{credit scoring}: before granting a loan, banks consider the borrower's income, age, financial history, collateral, ... to identify them as a high risk or low risk customer.

Credit scoring is a textbook machine learning problem which we will return to in Sec.\ \ref{secIVA}. Let us for now focus our attention on selecting the optimal features for credit scoring:  we want to determine which data on past applicants can provide useful information in determining the creditworthiness of new applicants. This problem arises when some of the data is irrelevant or weakly correlated to the output, when we do not have access to all the data, and when performing credit scoring by using all the data is too computationally expensive.

In Ref.\ \cite{Milne}, it was shown how this can be translated into a QUBO problem to be run on a quantum annealer. Let us define the matrix $U$; its columns represent the features of past credit applicant (e.g: age, etc), while its rows representing their numerical values. We also define a vector $V$, the record of past credit decisions. The cost function to be minimized is then: 
\beq
\label{eq:OptimalFeature}
w = - \left( \alpha \sum_{j = 1}^n x_j |\rho_{Vj}| - (1-\alpha) \sum_{j=1}^n \sum_{k\neq j}^n x_j x_k |\rho_{jk}| \right).  
\eeq
The binary boolean variable $x_i$ is $1$ if the feature $i$ is in the subset of ``selected'' features, and $0$ otherwise. The matrix $\rho_{ij}$ represents the correlation between columns $i$ and $j$ of matrix $U$, and vector $\rho_{Vj}$ represents the correlation between column $j$ of $U$ and the single column of $V$. The parameter $\alpha$ controls the relative weight between the two penalty terms. The first term models the influence that features have in the credit outcome, and the second models the independence of the features amongst themselves. The parameter $\alpha$ therefore controls the relative weight between the influence and the independence of the features. 

In this form, Eq.\ \eqref{eq:OptimalFeature} can be optimized by a quantum annealer. This was implemented as a proof-of-principle on the 1QBit SDK toolkit \cite{Milne}. These results show that future quantum annealers can also be used to determine optimal features in credit analysis. 

\section{Quantum Machine Learning}
\label{sec4}

The field of \emph{machine learning} broadly amounts to the design and implementation of algorithms that can be trained to perform a variety of tasks. These include pattern recognition, data classification, and many others. We call \emph{training} the process of optimizing the algorithm's parameters to recognize specific inputs (the training data). The trained algorithm can then be applied to assess new inputs. The field of classical machine learning has grown enormously, mainly due to hardware and algorithmic developments (allowing, for instance, to train deep learning networks) \cite{Alpaydin2010}. The basic principles of machine learning are at the root of a number of vastly successful fields, the most prominent of which is probably neural networks, which includes shallow networks, deep learning, recurrent networks, convolutional networks, and many more. Other machine learning approaches include principal component analysis, regressions, variational autoencoders, hidden Markov models, and more. 

We saw in Sec.\ \ref{secIIB} that machine learning is a key ingredient to tackle many financial problems. We also mentioned in Sec.\ \ref{QCforfinance} that a number of computing tasks could be run faster on a quantum computer. The purpose of this section is to bring these two ingredients together: we provide an overview of selected machine learning algorithms which are important to finance, and review developments which allow these algorithms to run faster on a quantum computer.

The field of QML is essentially separated in two main lines: the search for quantum versions of machine learning algorithms, and the application of classical machine learning to understand quantum systems. This set of ideas has recently emerged with a lot of momentum \cite{wittek2014,Schuld2014,Biamonte2016}.

Note that, while many QML algorithms are potentially ground breaking, many of them require the operation of a universal quantum computer. These are more advanced than quantum annealers, and are also more technologically challenging. In other words, while optimization problems can already benefit from the first generation of experimental quantum annealers, the implementation of certain QML algorithms won't be possible until the technology has developed further. At the current rate of experimental progress, however, we believe this will happen sooner rather than later.

\subsection{Data classification}
\label{secIVA}

Let us return to the problem of credit scoring which we first addressed in Sec.\ \ref{optcred}. The way this problem is typically dealt with relies on a machine learning method known as \emph{classification} \cite{Baesens2003}. Each data point (customer) is expressed as a vector, living in the vector space of all considered attributes (customer characteristics). The training set is labeled, such that each vector belongs to a class (the loan risk). When given a new vector, the program must determine the class it is most likely to belong to. One way to do this is by returning the class which occurs most frequently among the $k$ vectors which are closest to the new vector, with $k$ an integer.

% The program is trained with a dataset of past customers' characteristics, which is labeled to show which customers defaulted on their payments. The aim is to deduce a rule from these past records, such that, when the machine is provided with a new data point (customer), it can determine its label (the loan risk). 

Thus, we see that, in the case of credit scoring, classification algorithms are an essential tool for prediction. This area of machine learning is also at the heart of pattern recognition, which is extensively used for voice and facial recognition. Data classification algorithms are also used for outlier detection, where we identify points which are difficult to correctly attribute to a class. This type of process is essential for fraud detection \cite{Bolton2002}.

Depending on the size of the training set, and the number of attributes considered, finding a new vector's class can mean performing a large number of high dimensional projections. This can rapidly limit the confidence with which we can assign a class to the new vector, particularly for pattern recognition applications, where the number of attributes considered is gigantic. As a result, methods for running classification algorithms on a quantum computers generally focus on efficiently performing projection operations. 

% \begin{figure}
%  \centering
%  \includegraphics[width=0.98\linewidth]{figures/swap_test.pdf}
%  \caption{Swap test: the system is prepared in $\ket{0,a,b}$; H is the Hadamard operation, applied to the ancilla bit. This scheme relies on a controlled swap of $\ket{a}$ and $\ket{b}$, and results in a probability of measuring the ancilla bit in the ground state that is determined by the overlap between $\ket{a}$ and $\ket{b}$.}
%  \label{fig:swap_test}
% \end{figure}

A\"imeur, Brassard and Gambs pioneered the idea of recasting this problem on a quantum computer by expressing each data point as a quantum state \cite{Aimeur2006}. They build upon a proposal by Buhrman et al.\ \cite{Buhrman2001}, to efficiently estimate the classical distance $|\braket{a}{b}|$ between states $\ket{a}$, $\ket{b}$ by repeatedly performing swap tests. Lloyd et al.\ suggested an alternative method for encoding classical data into a quantum state. While their method also relies on performing a swap test, it boasts an efficiency which is superior to classical algorithms, even when the states' preparation is taken into account \cite{Lloyd2013a}.
% Similarly, a number of quantum data classification algorithms are developed in Ref.\ \cite{Ruan2017} which provide speedups relative to classical methods which range from polynomial to exponential.

Support vector machines are a subset of classification algorithms, and are among the most used supervised machine learning algorithms. These aim to find the hyperplane which separates a labeled dataset the most clearly in its two distinct classes. There exist a number of proposals to implement support vector machines on a quantum computer \cite{Rebentrost2014,Chatterjee2017}. These have attracted a lot of attention because the operations necessary to construct the hyperplane and assign a class to a new vector scale polynomially in $\log N$, where $N$ is the dimension of the vector space. This approach for implementing a support vector machine was demonstrated experimentally in Ref.\ \cite{Li2015a}.

% A fully quantum algorithm for pattern recognition was proposed by Trugenberger \cite{Trugenberger2002a}. This is done by initiating the system in a superposition of all training vector states. Given an input state to classify, a unitary evolution operator writes this state's Hamming distance to each training vector amplitude.

While this field is still in its infancy, there exist early suggestions to apply quantum classification algorithms to pattern recognition problems \cite{Trugenberger2002a,Schutzhold2003,Schuld2014,Ruan2017,Schuld2017}; pattern recognition on a quantum computer was recently demonstrated experimentally in Refs.\ \cite{Li2015a,Boyda}.

\subsection{Regression}

Another common financial problem is known as \emph{supply chain management}, which is the art of meeting customer demand while avoiding unwanted stock. As with classification algorithms, it can be essential to take into account a number of weak indicators when dealing with this type of problem. If we were, for instance, trying to estimate the number of umbrellas we are likely to sell in the following weeks, it would be essential for us to take a number of factors into account, including the weather forecast for this period.

This problem is qualitatively different from pattern recognition: given a new data point, instead of attributing a class to it, we want to learn a numeric function from the training data set. This process, known as \emph{regression}, is particularly useful when attempting an interpolation, and is consequently a core tool for economic forecasting \cite{Bontempi2013}. Regression algorithms are generally used to understand how the typical value of a response variable changes as an attribute is varied.

In most cases, the optimal fit parameters are found by minimizing the least-squares error between the training data and the values predicted by the model. This is usually done by finding the (pseudo)inverse of the training data matrix, a task which can be extremely computationally expensive for typical industrial datasets.

% \st{Problem: regression fails at forecasting when the factors affecting time series are not understood, or when the data is lacking. Additionally, the accuracy with which markets can be forecast decreases when the forecast result affects the behavior of the market.}

% \st{As with classification, it is the training which is most resource intensive. Quantum algorithms focus on finding optimal parameters with minimal resources \cite{Schuld2014}.}

There exists, however, a powerful linear algebra toolbox which allows us to diagonalize matrices on quantum computers exponentially faster than on classical computers \cite{Harrow2009}. The idea of applying this algorithm to perform regression on a quantum computer was pioneered by Wiebe, Braun, and Lloyd. They showed that, for a sparse training data matrix, they could encode the model's optimal fit parameters into the amplitudes of a quantum state. This can be done in a time which is exponentially faster than the fastest classical algorithm \cite{Wiebe2012}. Wang builds upon this work by applying modern methods for matrix inversion \cite{Childs2015}, and generalize this algorithm to non-sparse training data matrices  \cite{Wang2017}. 

A method for running an altogether different regression algorithm, known as Gaussian process regression, was suggested in Ref.\ \cite{Zhao2015}. This work also relies on the linear algebra toolbox provided by Ref.\ \cite{Harrow2009} to claim an exponential speedup relative to classical algorithms.

Wiebe et al.\ do note, however, that state tomography, which is necessary for the readout of these parameters, can be exponentially expensive in the data size. Schuld, Sinayskiy, and Petruccione circumvent this problem entirely by encoding their regression model into a quantum state, which is then used to perform predictions on new inputs directly \cite{Schuld2016}. As noted in Ref.\ \cite{Wang2017}, however, quantum states are delicate and difficult to store, making this method somewhat inconvenient.

\subsection{Principal component analysis}

In portfolio optimization, which we discussed in Sec.\ \ref{secIIB}, it is essential to have a global vision of interest rates paths, even when dealing with hundreds of swap instruments \cite{Darbyshire2016}. A standard tool for doing this is a machine learning algorithm known as principal component analysis (PCA).

The idea is simple: let $\vec{v}_j$ be a data vector of, for instance, changes in stock prices between times $t_j$ and $t_{j+1}$. We define the covariance matrix $C$ as $C \equiv \sum_j \vec{v}_j \vec{v}_j^T$, which encodes the correlations between the different stock prices at different times. The eigenvectors which correspond to small eigenvalues of $C$ (in absolute value) are called \emph{principal components}. These amount to the most important trends in the evolution of vectors $\vec{v}_j$, which we can use to predict future trends based on the history of the data. 

In practice, PCA amounts to finding dominant eigenvalues and eigenvectors of a very large matrix. Normally,  the cost is almost prohibitive. Indeed, usual algorithms for matrix diagonalization have a computational cost of $\mathcal{O}(N^2)$ for an $N \times N$ matrix, even if the matrix is sparse. For real-life data, where $N$ could be millions of stocks, this cost is simply astronomical.

It was recently shown that a version of this algorithm, the \emph{quantum-PCA algorithm}, could be run exponentially faster on a quantum processor \cite{Lloyd2014}. Specifically, this algorithm finds approximations to the principal components of a correlation matrix, with a computational cost of $\mathcal{O}((\log N)^2)$ (both in computational and query complexities). This development should vastly broaden the spectrum of applicability of PCA, allowing us to estimate risk and maximize profit in situations that were not feasible using classical methods. 

\subsection{Neural networks and associated models}

Neural networks have proved extremely successful at predicting markets and analyzing credit risk \cite{Trippi1992,Cristianini1999}. The key to this success lies in their ability to tackle tasks which require intuitive judgment and to draw conclusions even from incomplete datasets. These properties have made neural networks essential for, e.g., financial time-series prediction \cite{time, deep}. There exist a number of proposals to accelerate neural networks and deep learning algorithms through the power of quantum computing \cite{Adachi,Denil,Dumoulin}, which we will briefly review in the following. 

While machine learning algorithms are usually extremely efficient, their training can be computationally expensive. This overhead could be significantly reduced by training the neural network using a quantum annealer, such as the D-Wave or the Rigetti machines. Once trained, the algorithm can be run on any classical computer. We expect this method to be less prone to falling into local minima than standard training methods (such a gradient descent, Hessian methods, backpropagation, stochastic gradient descent, ...). 

In an early implementation of this idea, Ref.\ \cite{Benedetti2016} have shown that a Boltzmann machine could be efficiently trained using present day D-Wave quantum computer. This contribution was possible because neural networks do not require a general purpose quantum computer to run. {Boltzmann machines can be physically understood as classical Ising models where spin-spin couplings and local magnetic fields are fine-tuned, so that the thermal residual probability distribution of a subset of the spins mimics some input training probabilities.} While Boltzmann machines are not deep learning networks, we expect that this proof-of-principle study to be the first step in a number of truly ground breaking developments.

Alternatively, the training process could be sped-up exponentially (compared to classical training methods) by using quantum PCA methods to implement iterative gradient descent methods for network training  \cite{Lloyd2014}. Note that these approaches are generic: they could be applied to \emph{any} type of neural network, including shallow, convolutional, and recurrent networks.

Another possibility altogether is to design new, fully quantum neural network algorithms. This approach should allow the network to learn much more complex data patterns than those identifiable using a classical neural network. Early suggestions in this field include quantum perceptron models \cite{tab2} and quantum hidden Markov models \cite{qMar}. The latter is particularly relevant for us because (classical) hidden Markov models are commonly used for financial forecasting. Quantum hidden Markov models, being a generalization of their classical counterparts, promise richer dynamical processes. While promising, these ideas are still very much in their infancy, and need to be further studied before their power is fully understood.

\section{Quantum Amplitude Estimation and Monte Carlo} 
\label{sec5}

The Monte Carlo method is a a technique used to estimate a system's properties stochastically, by statistically sampling realizations of the system. It is ubiquitous in science, with applications in physics, chemistry, engineering, and finance. Where it really shines is in dealing with extremely large or complex systems, which cannot be approached analytically or handled through other methods.

In finance, the stochastic approach is typically used to simulate the effect of uncertainties affecting the financial object in question, which could be a stock, a portfolio, or an option. This makes Monte Carlo methods applicable to portfolio evaluation, personal finance planning, risk evaluation, and derivatives pricing \cite{Wilmott}. 

Imagine we want to sample a probability distribution which has width $\sigma^2$ and mean $\mu$. {The weak law of large numbers, which follows from Chebyshev inequality, tells us that it is sufficient to take $k = \mathcal{O}(\sigma^2 / \epsilon^2)$ samples with a prefactor of $\approx 10$} to estimate $\mu$ with approximately $99 \%$ probability of success. Mathematically, this can be expressed as:
\beq
\label{Chebyshev}
{\rm Pr} (|\tilde{\mu} - \mu| \ge \epsilon) \le \frac{\sigma^2}{k \epsilon^2}, 
\eeq
where $\epsilon$ is the error and $\tilde{\mu}$ is the approximation to $\mu$ obtained from $k$ samples. {In other words, the ratio $\sigma^2 / \epsilon^2$ dictates the speed of convergence with the number of samples $k$.} If we want to obtain the most probable outcome of a wide distribution, or obtain a result with a very small associated error, the required number of Monte Carlo simulations can become gigantic. This is the case for stock market simulations, for instance, which are routinely day-long simulations.

In this situation, obtaining a quantum speedup could make a notable difference. The first steps in this direction were done by Brassard, Hoyer, Mosca and Tapp in Ref.\ \cite{qaest}. In the first part of their paper, they extend Grover's search algorithm \cite{Grover1996} to construct an algorithm for Quantum Amplitude Amplification (QAA). Given a desired state with probability amplitude $p$, Brassard et al.\ show that they can amplify this probability to almost one in $\mathcal{O}(1/\sqrt{p})$ operations, i.e: quadratically faster than the best possible classical algorithm.

In the second part of their paper, Brassard et al.\ derive their Quantum
Amplitude Estimation (QAE) algorithm. QAE is an integral part of a large number of more complex quantum algorithms. In particular, it can be used to obtain a quadratic speed-up in the calculation of expectation values by Monte Carlo sampling. We shall return to this point shortly.

The QAE algorithm applies a series of QAA operations, followed by a QFT from Shor's quantum factoring algorithm \cite{qft}, to measure the approximate amplitude of any given state $| \Psi \rangle$. Specifically, if $| \Psi \rangle$ has a probability amplitude $p$, then $p$ can be estimated in $M$ calls to the oracle, with an error $\epsilon = 2 \pi \sqrt{p(1-p)}/M + \pi^2 / M$, and a probability $\geq 8/\pi^2$ of the measure being successful. Discussing the technical details of this algorithm's implementation are beyond the scope of this paper. We refer the interested reader to Ref.\ \cite{qaest} and Chapter 6 of Ref.\ \cite{Nielsen2010} for more details. 

Building upon this result, Montanaro showed that Monte Carlo simulations can run on a quantum computer, obtaining the same accuracy as predicted by Eq.\ \eqref{Chebyshev} but with almost quadratically less samples $k$ \cite{Montanaro2015}. Specifically, Montanaro's algorithm requires only $\mathcal{O}(\sigma / \epsilon)$ samples (up to polylogarithmic factors) to estimate $\mu$ with $99 \%$ success probability. The source of this speedup is the QAE algorithm, which is at the heart of the efficient estimation of the distribution's mean.

Note that the process of sampling the random variable distribution can either be a classical or a quantum process. If the random sampling is performed through a quantum algorithm, this can lead to a speedup. When used in combination with the quantum algorithm for sampling described above, then both speedups \emph{concatenate}.

In the following, we review two recent articles which suggest to apply quantum-accelerated Monte Carlo to problems in finance. 

\subsection{Pricing of financial derivatives} 

Let us now return to the problem of financial derivatives, which we first mentioned in Sec.\ \ref{FinanceCoreProblems}. These contracts have a payoff that depends on the future price trajectory of some asset, which may have a stochastic nature. Brokers must know how to assign a fair price to the derivatives from the state of the market. This is the \emph{pricing} problem. The classical approach to this problem is via simplified scenarios, such as the Black-Scholes-Merton model \cite{bsm, bsm2} and Monte Carlo sampling.

Building upon Montanaro's work, Rebentrost, Gupt, and Bromley suggested using quantum-accelerated Monte Carlo to obtain a quadratic speedup in pricing financial derivatives \cite{Rebentrost2018}. The idea is to design a quantum operator which has the same probability distribution as the financial derivative, and apply the method from Ref.\ \cite{Montanaro2015} to estimate its expectation value. 
Rebentrost et al.\ also discuss how their method can be applied to the pricing of a European call option and to Asian options \cite{Rebentrost2018}. 

\subsection{Risk analysis}

Financial institutions need to be able to accurately manage and compute risk, which we introduced in Sec.\ \ref{FinanceCoreProblems}. One way to mathematically quantify the risk is through the Value at Risk (VaR) function, which measures the distribution of losses of a portfolio. For a given probability distribution, VaR$_\alpha(X)$ is defined as the smallest value $x \in [0, N-1]$ such that ${\rm Pr}(X \le x) \ge (1- \alpha)$, where $\alpha$ is a confidence level, $\alpha \in [0, 1]$. Another useful risk estimation tool is the Conditional Value at Risk (CVaR), which measures the expected loss of a portfolio for losses greater than VaR.

Typically in quantitative finance, VaR and CVaR are estimated using Monte Carlo sampling of the relevant probability distribution. Woerner and Egger pioneered the idea of adapting the core principles of quantum-accelerated Monte Carlo to efficiently estimate these variables \cite{Woerner2018}. Specifically, by applying the QAE algorithm, which called a tailored oracle function, they were able to determine VaR and CVaR with excellent accuracy and a quadratic speedup relative to classical methods. The authors of Ref.\ \cite{Woerner2018} went as far as constructing an implementation of their algorithm as a quantum score, which was tested for some small examples on the IBM Q Experience. This is interesting because small-size experiments such as this one could already show a quantum speed-up relative to classical methods.

\section{Conclusions and perspectives}
\label{sec7} 

In this paper, we have reviewed ways in which quantum computing could disrupt finance. As we have seen, this field is developing at a striking rate, partly due to experimental developments in quantum hardware, which are surpassing all expectations, and partly due to conceptual leaps, which promise gigantic speedups for widely applicable algorithms. It is our belief that before long quantum computers will play a key role in quantitative finance. 

We caution the reader that a number of experimental breakthroughs will be necessary before we can construct a universal quantum processor capable of surpassing present-day supercomputers. We will need, for instance, to vastly increase the quality of qubits to implement some of the algorithms detailed here. It is possible, however, that faulty quantum computers will find interesting applications far before we achieve fault-tolerant quantum computing. We would expect it is in this area that the first real disruptions to finance will occur, and urge researchers to investigate this fascinating direction of study.

There are a number of applications that quantum physics finds in economy that, while fascinating, we chose not to cover. It would have been interesting to talk about how quantum technologies can be relevant to the blockchain and cryptocurrencies \cite{bitcoin}, or to discuss quantum finance \cite{Haven2002,baaquie2007}, quantum money \cite{Wiesner1983}, the impact of quantum cryptography in the security of financial transactions \cite{bb84, Eckert91}, {and the applications of quantum simulators \cite{qs1, qs2} in finance.} These could constitute an interesting subject for a future publication.

\bigskip 
\emph{Acknowledgements.-} We acknowledge discussions with  A. Cadarso, J. I. Latorre, D. Marcos, M. Masoliver, and A. Rubio-Manzanares.

\bibliographystyle{apsrev4-1_our_style}
\bibliography{bibliography.bib}

\end{document}